\documentclass{JHEP}
\input epsf.tex
\epsfclipon

\usepackage{epsfig}
\usepackage{epstopdf}
\usepackage{epsf}
\input{epsf.sty}
\usepackage{graphicx,amsmath,amssymb}
\usepackage{epsfig,multicol}

\usepackage{bbm,bm,amsmath,amssymb}

\def\be{\begin{equation}}
\def\ee{\end{equation}}
\def\baray{\begin{eqnarray}}
\def\earay{\end{eqnarray}}

\newcommand{\roughly}[1]{\mathrel{\raise.3ex\hbox{$#1$\kern-0.85em
\lower1ex\hbox{$\sim$}}}}

\def\2pi{\left(2\pi\right)}

\def\beq{\begin{equation}}
\def\eeq{\end{equation}}
\def\beqa{\begin{eqnarray}}
\def\eeqa{\end{eqnarray}}
\def\bea{\begin{eqnarray}}
\def\eea{\end{eqnarray}}
\def\sfrac#1#2{\textstyle\frac#1#2}
\def\sss{\scriptscriptstyle}
\def\D3{\overline{\mbox{D3}}}

\def\exd{{\rm d}}
\def\pref#1{(\ref{#1})}
\def\nn{{\nonumber}}

\def\cD{{\cal D}}
\def\cF{{\cal F}}
\def\cG{{\cal G}}
\def\cH{{\cal H}}

\def\cT{{\cal T}}

\title{Uplifting and Inflation with D3 Branes}

\author{C.P.\ Burgess,$^{1,2}$ James M.\ Cline,$^3$
Keshav Dasgupta,$^3$ Hassan Firouzjahi$^{3}$\\
${}^1$ Dept. of Physics \& Astronomy, McMaster University\\
\qquad 1280 Main St. W, Hamilton, Ontario, Canada L8S 4M1. \\
${}^2$ Perimeter Institute for Theoretical Physics\\
\qquad 31 Caroline Street N, Waterloo, Ontario, Canada N2L 2Y5. \\
$^3$Physics Department, McGill University, 3600 University
Street,\\ \qquad Montr{\'e}al,  Qu{\'e}bec, Canada H3A 2T8 }
\date{1 October 2006}

\abstract{Back-reaction effects can modify the dynamics of mobile
D3 branes moving within type IIB vacua, in a way which has
recently become calculable. We identify some of the ways these
effects can alter inflationary scenarios, with the following three
results: ($1$) By examining how the forces on the brane due to
moduli-stabilizing interactions modify the angular motion of D3
branes moving in Klebanov-Strassler type throats, we show how
previous slow-roll analyses can remain unchanged for some brane
trajectories, while being modified for other trajectories. These
forces cause the D3 brane to sink to the bottom of the throat
except in a narrow region close to the D7 brane, and do not
ameliorate the $\eta$-problem of slow roll inflation
in these throats; ($2$) We argue that a recently-proposed
back-reaction on the dilaton field can be used to provide an
alternative way of uplifting these compactifications to Minkowski
or De Sitter vacua, without the need for a supersymmetry-breaking
anti-D3 brane; and ($3$) by including also the $D$-term forces which
arise when supersymmetry-breaking fluxes are included on D7 branes
we identify the 4D supergravity interactions which capture the
dynamics of D3 motion in D3/D7 inflationary scenarios. The form of
these potentials sheds some light on recent discussions of how
symmetries constrain $D$ term interactions in the low-energy
theory.}

\maketitle

\begin{document}

\section{Introduction}

Of late there has been considerable interest in understanding the
dynamics of time-dependent string configurations largely motivated
by their possible applications to cosmology, particularly to the
search for inflation. The resulting dynamics has begun to produce
a number of interesting scenarios for inflationary constructions,
such as the brane-antibrane mechanism
\cite{DvaliTye}-\cite{realistic}, D3/D7 models \cite{DHHK, D3D7},
modular inflation \cite{Modular} and other scenarios
\cite{dbi}-\cite{Ohta}.

The study of the motion of branes within type IIB geometries with
fluxes has been of particular interest, given the progress towards
modulus stabilization which these geometries provide
\cite{GKP}-\cite{DDF}. The goal of these studies is to improve the
control over the approximations being made, in an effort to more
reliably tie the time-dependent behaviour to the properties of
real string vacua. What makes these studies difficult is the
necessity of exploring nonsupersymmetric configurations when
exploring time-dependent solutions, with the loss of control over
the corrections to the leading features that this supersymmetry
breaking entails.

The greater calculational control offered by supersymmetry makes
it advantageous to keep any supersymmetry breaking parameterically
small, putting a premium on constructions which can be regarded as
only small deviations from the supersymmetric limit. In
particular, this makes it preferable not to break supersymmetry
with antibranes, since these necessarily require that
supersymmetry be nonlinearly realized.\footnote{Or explicitly
broken, which amounts to the same thing for gauge symmetries
\cite{BL}.} This motivates identifying dynamical situations where
supersymmetry breaks spontaneously, but in a way which allows a
description of the low-energy dynamics purely in terms of a 4D
supergravity. The motion of a mobile, or itinerant, D3 brane
moving in the presence of background fluxes and the fields
generated by various D7 and O7 sources provides an attractive
class of systems of this type.

In this paper we use recent calculations \cite{BDKMMM,BHK} of the
back-reaction of such an itinerant D3 brane on the low-energy
gauge coupling functions to identify some of the forces acting on
the mobile D3. The first of these forces is described in the
low-energy theory by an $F$-term potential \cite{BDKMMM}, which
arises because the low-energy superpotential acquires a dependence
on the low-energy gauge coupling function due to nonperturbative
effects. These can arise either through Euclidean D3-branes
wrapping a 4-cycle of the Calabi-Yau (CY) \cite{Witten}, or by
gaugino condensation \cite{GC} of an unbroken gauge group living
on a stack of D7 branes which also wrap a 4-cycle in the CY. The
potential induced in this way was identified some time ago
\cite{KKLMMT} as the source of an $\eta$ problem for inflationary
models involving mobile D3 branes moving in the throat; ignorance
of its detailed form left open the possibility \cite{KKLMMT} --
\cite{realistic} that inflation could  nonetheless be achieved by
adjusting inflaton-dependent corrections to the superpotential.
But this begged the question of whether string theory really
provides the desired kind of superpotential corrections.

Progress toward computing this superpotential more explicitly was
recently made by ref.\ \cite{BDKMMM}, who noted that the form of
the superpotential corrections can be explicitly calculated for
branes moving within a Klebanov-Strassler (KS) throat
\cite{KW,KS}, if the wrapped 4-cycle is itself within the throat.
In section \S\ref{sec D3 in warped throats}  we use this
calculation to compute the motion of a mobile D3 brane, and find
that the resulting superpotential vanishes when minimized along
the angular directions which parameterize the 5D $T^{1,1}$
submanifold of the throat. We conclude that the superpotential by
itself therefore does not counteract the $\eta$ problem for motion
along these directions. Furthermore, the forces which push the D3
brane in the angular directions are too steep to provide new
slow-roll mechanisms themselves.

An additional force on the D3 brane also arises at the same order
as the superpotential correction, due to the back-reaction of the
D7 brane on the background dilaton profile \cite{Ouyang}. We find
in \S\ref{sec: uplifting} that this force competes with that due
to the superpotential in such a way as to change the stable point
of the D3 motion in the angular directions in the throat. However
including these corrections does not allow one to fine-tune away
the $\eta$-problem of slow roll inflation, since they provide a
force on the D3 brane which is in the same direction as the force
which gives rise to the $\eta$ problem. On the other hand, we show
that because the dilaton correction increases the potential energy
of the D3,  it can be used to uplift the vacuum to de Sitter or
Minkowski space, even in the absence of an antibrane. This new
possibility may allow better control over the corrections to this
picture inasmuch as it does not rely on the introduction of
badly-broken supersymmetry via antibranes.

Finally, in section \S\ref{sec: dterm} we use the modified gauge
kinetic functions of ref.~\cite{BDKMMM} to compute the force on a
mobile D3 brane which arises when supersymmetry-breaking fluxes
are introduced on the D7 branes, due to the partial failure of the
BPS cancellations amongst the interbrane forces. This type of
force has been argued to be described by a $D$-term potential
\cite{Dterm}-\cite{BKQ}, whose form has also been recently
identified by other workers \cite{HKLVZ}. We identify the $F$ and
$D$ terms which describe the motion of a D3 in the special case
that the internal dimensions have the form of a product of a
4-cycle with two toroidal (or orbifold) dimensions (such as for
$K3 \times T^2/\mathbb{Z}_2$). Besides expecting these results to
be pertinent for constructing new inflationary scenarios, we find
they also provide useful illustrations of how the low-energy
theory implements some of the symmetries which arise when
discussing $D$-term potentials.

\subsection{D3's and D7's in type IIB vacua}

In type IIB compactifications the perturbations to the strength of
gauge couplings on nearby D7 branes due to the presence of
itinerant D3 branes within the bulk have recently been calculated
in two independent ways. They were first obtained as
loop-generated threshold effects within an open-string picture
\cite{BHK}, with the result in some instances reconfirmed by
performing a back-reaction calculation within the closed-string
picture \cite{BDKMMM}.

The background spacetimes which arise within warped type IIB
compactifications preserving $N=1$ supersymmetry in four
dimensions \cite{GKP,sethi} have the general form
\be \label{metric}
    \exd s^2 = h^{-1/2}(y) \, g_{\mu\nu}(x) \, \exd x^\mu \exd x^\nu
    + h^{1/2}(y) \, \tilde g_{mn}(y) \, \exd y^m \exd y^n
    \,,
\ee
where $h(y)$ denotes the warp factor. Given this metric, the gauge
term for a massless gauge field situated on a stack of coincident
space-filling D7 brane becomes
\be
    S_g = - \frac14 \int_\Sigma d^4x \sqrt{-g_4} \;
    g^{\mu\nu} g^{\lambda\rho} \cG_{ab} F^a_{\mu\lambda}
    F^b_{\nu\rho} \,,
\ee
where the matrix $\cG_{ab}$ governs the effective gauge
coupling strength, and is given semiclassically by the following
warped volume
\be
    \cG_{ab} = \cT_7 V_\Sigma^w \delta_{ab} \equiv \cT_7 \,
    \delta_{ab} \int_\Sigma d^4y \sqrt{\tilde g_4}
    \; h \,,
\ee
where $a$, $b$ label gauge group generators, $\Sigma$ denotes the
4-cycle wrapped by the 7-brane in the internal 6 dimensions and
$\cT_7 \propto \alpha'^2 T_7$ is proportional to the 7-brane
tension.

For compactifications which preserve $N=1$ supersymmetry in 4D,
supersymmetry dictates that $\cG_{ab}$ must be the real part of a
holomorphic quantity when it is viewed as a function of the moduli
which appear as fields within the low-energy 4D theory. For
instance, if we follow in this way the dependence on the volume
modulus for the internal dimensions, $\tilde g_{mn} = e^{2u} \,
g_{mn}$, we see that $\cG_{ab} \propto e^{4u} \, \delta_{ab}$. In
the absence of the itinerant D3 branes of interest below, this is
indeed the real part of a holomorphic function because it is
related to the holomorphic volume modulus, $\rho$, by
\be
    e^{4u} = \rho + \overline\rho \,.
\ee

\subsubsection*{D3 perturbations to D7 gauge couplings}

When identifying forces acting on the D3's our interest is in how
the above expressions respond to the presence of a perturbing
mobile D3 brane situated at a point $y^m = z^m$ in the bulk, since
we wish to follow how the low-energy 4D theory depends on the
position modulus, $z^m$. From the microscopic point of view, this
involves determining how the the quantity $\cG_{ab}$ responds to
the back-reaction of the 10D geometry to the 3-brane's position.
$\cG_{ab}$ depends on this back-reaction because the gauge
couplings depend on the volume of the cycle which the D7 wraps,
and this becomes corrected by the change in the bulk geometry due
to the presence of the itinerant D3 brane.

As was recently emphasized \cite{BDKMMM}, the holomorphy of this
back-reaction enters in two ways: through perturbations to the
volume, $e^{4u}$, and to perturbations to $h$. Although both of
these perturbations introduce nonholomorphic contributions to
$\cG_{ab}$, these contributions cancel to leave a holomorphic
brane-dependent contribution to the gauge kinetic function:
$\cG_{ab} \propto \delta_{ab} \, \hbox{Re}\, f$ with $f = \rho +
F(z)$, in agreement with the earlier arguments of \cite{BHK}. The
quantity $F(z)$ here represents a suitable cycle average of the
holomorphic part of the appropriate Greens function which governs
the perturbations $\delta e^{4u}$ and $\delta h$.

\subsection{Forces on itinerant D3 branes}

Although all static forces cancel (by definition) between an
itinerant D3 brane and the other branes in a lowest-order
supersymmetric construction, typically this no longer remains true
once all corrections are taken into account. For sufficiently slow
motion the dynamics generated by these forces can be described
within a low-energy effective 4D theory, whose comparative
simplicity is often a prerequisite for being able to fully analyze
the motion. Furthermore, provided that the size of the
supersymmetry breaking associated with the forces is kept small
enough the result must be a particular kind of $N=1$ 4D
supergravity. An approximately supersymmetric 4D limit of this
type is particularly powerful because of the control it provides
over the approximations underlying the compactification which
leads to the effective 4D Lagrangian.

The key to understanding the 4D supergravity formulation lies with
the gauge kinetic function given in the previous section, which
have the generic form $\cG^s_{ab} = \cT_7 V_\Sigma^i \,
\delta_{ab} \left(f_s + \overline{f_s} \right)$ with
\be \label{eq: kinfn}
    f_s(\rho,z) = \rho + F_s(z)  \,.
\ee
Here the subscript `$s$' labels the relevant D7 brane on which
lives the corresponding massless gauge field. This dependence of
gauge couplings on D3 brane positions gives rise to two kinds of
forces of this type which are of particular interest.

\subsubsection*{$F$-terms and volume-stabilization forces}

There are forces on the D3 brane arising from any dynamics (like
gaugino condensation or D3 instantons) which stabilize the various
moduli of the bulk geometry. These mediate a force on the
itinerant D3 because the moduli typically adjust in the
supersymmetric case as the D3 moves, as required in order to keep
the net force on the D3 zero. Any energetic cost for making this
adjustment indirectly induces a potential which depends on the D3
position. This leads to the force which so complicated the
inflationary analysis within warped throats
\cite{KKLMMT,realistic}

The interaction potential generated in 4D by modulus stabilization
is incorporated by using the kinetic function, eq.~\pref{eq:
kinfn}, within the standard superpotentials which are used in the
literature to describe the gaugino condensation (or D3 instanton)
which stabilizes the relevant moduli \cite{KKLT,BDKMMM,GC}:
\be \label{WForm}
    W = W_0 + \sum_s A_s \exp \Bigl[ -a_s f_s(\rho,z) \Bigr] \,.
\ee
Here the constant $W_0$ expresses the effects of any
supersymmetry-breaking amongst the higher-dimensional fluxes which
stabilize some of the moduli, while the exponential term contains
the influence of gaugino condensation (or the like) on various D7
branes, and involves the dependence on the D3 position due to the
back-reaction of the D3 onto the relevant gauge coupling
strengths. The quantities $A_s$ and $a_s$ are $z$-independent
constants which are calculable given the details of the underlying
physics.

\subsubsection*{$D$-terms and direct interbrane forces}

A supersymmetry-breaking flux localized on any of the branes also
introduces a direct interbrane force, independent of the
stabilization of bulk moduli. It does so because the breaking of
supersymmetry ruins the BPS cancellations among the long-range
bulk forces corresponding to the exchange of massless
closed-string states. This typically leads to an interbrane
potential which varies near a source brane like $r^{2-d}$, where
$d$ counts the number of transverse dimensions. (For $d=2$ the
potential becomes logarithmic.)  It gives rise to the attractive
force which is used in most brane-antibrane \cite{BBbar} and D3/D7
\cite{DHHK,D3D7} inflationary analyses.

A similar argument can be used to incorporate the effects of
imperfect cancellation of bulk forces due to
supersymmetry-breaking fluxes localized on various branes. To this
end, imagine turning on a background gauge field, $\cF_{mn}$, on a
7-brane, which at the classical level contributes a 4D
Einstein-frame action of the form
\be
    S_f = - \cT_7 \int d^{\,4}x \sqrt{g_4} \; \cH \,,
\ee
where
\be
    \cH = \int_\Sigma d^4y \sqrt{g_4} \; e^{-12u}
    h^{-1} g^{mn} g^{pq} \cF_{mp} \cF_{nq}
\ee

If the fluxes involved break supersymmetry by a sufficiently small
amount, then it has been argued \cite{Dterm}-\cite{BKQ} that such
a flux term contributes a $D$-term potential to the low-energy
effective 4D supergravity, of the form
\be \label{VDterm}
    V_D = \frac12 \, \cG^{ab} D_a D_b \,,
\ee
where $\cG^{ab}$ is the inverse of the matrix, $\cG_{ab}$, of
holomorphic gauge-kinetic functions, and the sum is over the
generators of the group gauged by the massless spin-1 particles.
Here
\be \label{DForm}
    D_a = \xi_a + {\cal D}_a
\ee
where ${\cal D}_a = \partial_i K(\varphi,\overline\varphi) [t_a
\varphi]^i$ represents the $D$-term contribution of any charged
scalar matter fields in the low-energy theory, and $\xi_a$
represents a field-dependent Fayet-Iliopoulos (FI) term \cite{FI},
which can only arise for $U(1)$ factors of the gauge group.

\section{D3 dynamics in warped throats}
\label{sec D3 in warped throats}

We next turn to a more detailed exploration of the implications of
these forces for the specific case of D3 motion in the
strongly warped region of the internal geometry. The internal geometry
will turn out to be more complicated than the simple deformed conifold
case studied earlier. In fact the susy breaking for our case will be
intimately tied up with some specific features of the internal geometry.
In the following section we will elaborate this story.

\subsection{Internal geometry and supersymmetry breaking}

We therefore begin by analysing the precise background metric for
our case. In terms of eq.~\pref{metric} we aim to determine
$\tilde g_{mn}$ that would include the backreactions from the D7
branes, D3 branes and fluxes, as well as some controlled
nonperturbative effects such as gaugino condensates on the D7
branes. The warp factor $h$ appearing in \pref{metric} already
takes into account some of these effects, but there are in
addition some subtle changes to the standard deformed conifold
background that are responsible for breaking supersymmetry in our
case. In fact we soon argue that even after we switch off the
nonperturbative SUSY-breaking effects, the resulting background
still breaks supersymmetry because the backreactions from the D7
and the D3 branes do not allow {\it primitive} three-form fluxes
in this geometry.

The three-form fluxes are the $H_3 = H_{NS} $ and $F_3 = H_{RR}$
which would satisfy the equations of motion if $G_3 \equiv F_3 +
\tau H_3$ is imaginary self-dual (ISD) where $\tau$ is the
axio-dilaton. We can view the RR component of $G_3$ as coming from
a dual theory with wrapped D5 branes.

The metric now can be computed using various arguments. In the
absence of itinerant D3 branes, the background with D7 branes and
fluxes could be supersymmetric. Ouyang \cite{Ouyang} has computed
the metric of D7 branes with fluxes on a Klebanov-Tseytlin type
geometry \cite{KlebT} by regarding the D7 branes as probes in the
background. The final metric therein has an overall warp factor
$h$ (much like \pref{metric}) with $\tilde g_{mn}$ being given by
the Klebanov-Tseytlin metric \cite{KlebT}. However there are two
shortcomings: one, the geometry should be given by a full F-theory
picture, and two, the metric $\tilde g_{mn}$ should be the
Klebanov-Strassler type metric \cite{KS}. These shortcomings are
not very severe as long as we are away from the tip of the
Klebanov-Strassler throat, and consider the geometry only in the
neighborhood of one D7 brane. On the other hand, once we try to
incorporate all the branes the global geometry becomes very
complicated and we can only infer the local picture in certain
cases \cite{GT}.

Here we try to address the noncompact limit of our global geometry
when additional D3 branes are also incorporated into the picture.
From the considerations of \cite{GT} and \cite{Minasian} (see also
\cite{Dymarsky}) the metric of the internal space looks like:
\beq \label{fimet}
    \exd s^2 = F_1 ~(e_1^2 + e_2^2) + F_2 \sum_{i =
    1}^2~(\epsilon_i^2 - 2be_i \epsilon_i) + v^{-1}(\epsilon_3^2 +
    \exd r^2) \,,
\eeq
where $F_i$ are some functions of the radial coordinate $r$, ($v,
b$) are parameters, and ($e_i, \epsilon_i$) are defined as
\beqa \label{eidef}
    e_1 &=& \exd\theta_1,~~~~~~e_2 = -{\rm sin}~\theta_1 \exd\phi_1 \,,
    \nonumber\\
    \epsilon_1 &=& {\rm sin}~\psi~{\rm sin}~\theta_2 \exd\phi_2
    + {\rm cos}~\psi~\exd\theta_2 \,,
    \nonumber\\
    \epsilon_2 &=& {\rm cos}~\psi~{\rm sin}~\theta_2 \exd\phi_2
    +  {\rm sin}~\psi~\exd\theta_2 \,,
    \nonumber\\
    \epsilon_3 &=& \exd\psi + {\rm cos}~\theta_1 \exd\phi_1
    + {\rm cos}~\theta_2 \exd\phi_2 \,.
\eeqa
We see that the background is neither Klebanov-Tseytlin nor
Klebanov-Strassler because there are two two-spheres -- given by
$\sum_i e_i^2$ and $\sum_i \epsilon_i^2$ -- that have unequal
radii. In fact the radii are given by $F_1(r)$ and $F_2(r)$
respectively. Once we remove the D7 branes, the $F_i$ take the
following form \cite{PaT}
\beq \label{ptsol}
    F_1(r) = e^g + b^2\, e^{-g}, ~~~~~~~~ F_2(r) = e^{-g} \,,
\eeq
where $g = g(r)$ whose functional form can be extracted from
\cite{PaT}, and $b$ measures the size of the three-cycle at the
tip of the throat. On the other hand, if we remove the D3 branes
and put in a single D7 brane, then the background will be close to
the one predicted by Ouyang \cite{Ouyang} with $F_1 \approx F_2$
but with a naked singularity at the tip. If we replace the naked
singularity with a nontrivial three-cycle then the local geometry
is given in \cite{GT} with $F_1 \ne F_2$. Thus incorporating all
the branes and fluxes, the geometry around the neighborhood of the
D3 and the D7 branes will clearly be \pref{fimet} with $F_1 \ne
F_2$.

To study the issue of supersymmetry, let us take the limit $b \to
0$. In this limit our background looks like a warped resolved
conifold with $G_3$ fluxes and branes. Supersymmetry will be
preserved if we can argue that $G_3$ is a {\it pure} (2,1) form
with vanishing (1,2), (3,0) and (0,3) components. Recall that this
condition is stronger than the ISD condition imposed on the $G_3$
fluxes.

The background supergravity equations of motion connect these forms
to radii of various cycles in the internal manifold \pref{fimet}. For
example,
a direct computation of background fluxes along the lines of
\cite{cvetic} reveals that $G_3$ has both (2,1) as well as
(1,2) components given in the following way:
\beq \label{twne}
    G_3 ~ = ~ (F_1^{-1} + F_2^{-1})\Lambda^{2,1} ~
    \oplus ~ (F_1^{-1} - F_2^{-1}) \Lambda^{1,2} \,,
\eeq
and would become supersymmetric when $F_1 = F_2$ globally.
Here we have denoted the (2,1) and (1,2) forms by $\Lambda^{2,1}$ and
$\Lambda^{1,2}$ respectively.
Locally
imposing this cancellation implies that the regime of interest has
$G_3 \propto \Lambda^{2,1}$ with vanishing (1,2) piece. However it
would be wrong to conclude that supersymmetry is restored! Only
under strict global cancellation of the (1,2) part could we infer
unbroken supersymmetry. From the analysis presented here (along
with the earlier references) it seems difficult to have $F_1 =
F_2$ everywhere in the internal space.

Consider now the case when $F_1 \ne F_2$ but $H_3 = 0$ and $F_3 \ne 0$.
This could in principle happen when we are in the far-IR region of the geometry.
Now of course the concept of (2,1) and (1,2) forms makes no sense as $G_3$ is
real. Is SUSY restored for our case? The answer is no because in the presence of
$F_3$ and $F_1 \ne F_2$, a D7 brane breaks all supersymmetry. Therefore we see that
SUSY could in principle be broken by two effects in the background \pref{fimet}:
non-primitive $G_3$ fluxes and D7 branes. Both these effects are somehow connected to
having $F_1 \ne F_2$.

\subsection{The superpotential correction}

We now return to our starting point, the warped solution,
eq.~\pref{metric}, where $\exd s_6^2$ given by the metric
\pref{fimet} and $h(r)$ is the warp factor. To simplify the
ensuing analysis, let us first consider the following limits of the
variable defined in \pref{fimet}:
\beq \label{limiting}
    b~\to~0, ~~~~v ~=~1, ~~~~ F_1(r) ~ \approx ~F_2(r)~ =
    ~ {r^2\over 6}, ~~~~~\epsilon_3 ~ \to ~{r\epsilon_3\over 3} \,.
\eeq
which gives the conifold limit of the metric.
We employ the standard idea that this manifold can be
defined by the complex surface
\beq \label{cone}
    w_1\,  w_2- w_3 \, w_4=0 \,,
\eeq
in four complex dimensions. The complex coordinates $w_i$ are
related to real coordinates $(r, \theta_1, \theta_2, \phi_1,
\phi_2, \psi)$ via
\beqa \label{wdef}
    w_1&=&r^{3/2}\, e^{\frac{i}{2} (\psi-\phi_1-\phi_2)} \sin
    \frac{\theta_1}{2} \, \sin \frac{\theta_2}{2} \, ,
    \nonumber\\
    w_2&=&r^{3/2}\,  e^{\frac{i}{2} (\psi+\phi_1+\phi_2)} \cos
    \frac{\theta_1}{2} \, \cos \frac{\theta_2}{2}  \, ,
    \nonumber\\
    w_3&=&r^{3/2}\,  e^{\frac{i}{2} (\psi+\phi_1-\phi_2)} \cos
    \frac{\theta_1}{2} \, \sin \frac{\theta_2}{2}  \, ,
    \nonumber\\
    w_4&=&r^{3/2}\,  e^{\frac{i}{2} (\psi-\phi_1+\phi_2)} \sin
    \frac{\theta_1}{2} \, \cos \frac{\theta_2}{2}  \,,
\eeqa
In terms of these real coordinates, the metric of the conifold can
be written explicitly,
\beq \label{conemetric}
    \exd s_6^2= \exd r^2 + r^2 \left[
    \frac{1}{9} \left(\exd \psi +\sum_{i=1}^2 \cos \theta_i \, \exd \phi_i\right)^2 +
    \frac{1}{6} \sum_{i=1}^2 \left( \exd \theta_i^2 +\sin^2 \theta_i \, \exd \phi_i^2 \right)
    \right] \,.
\eeq
As emphasised before, the limit \pref{limiting} provides a good
approximation only for intermediate ranges within a throat for two
reasons. First, in this regime $G_3$ is a pure (2,1) form and the
complications due to the other components do not show up; and
secondly our results do not depend on these complications
associated with $b \ne 0$, since all the action will be taking
place away from the bottom of the throat.

The K\"ahler potential for the moduli of such a configuration is
known to be \cite{DG}
\beq \label{Kahlerform}
    \kappa_4^2 K 
    = -3\log\left[ 2\sigma - c\, k(w_i,\bar w_i)\right]  \,,
\eeq
%
where $c = \frac13 \kappa_4^2 T_3$ and  $2 \sigma = \rho +
\bar\rho$. The quantity $k$ denotes the K\"ahler potential of the
Calabi-Yau space itself --- in our case, the conifold --- which is
given by \cite{CO,HKN}
\beq \label{kpcy}
    k(w_i,\bar w_i) = r^2 = \left(\sum_{i=1}^4 |w_i|^2\right)^{2/3}\, .
\eeq
This form can be inferred in various ways, for example by
comparing the D3-brane kinetic term in the present SUGRA
description with the expression that results from evaluating the
DBI action in the compactified background.

\EPSFIGURE[ht]{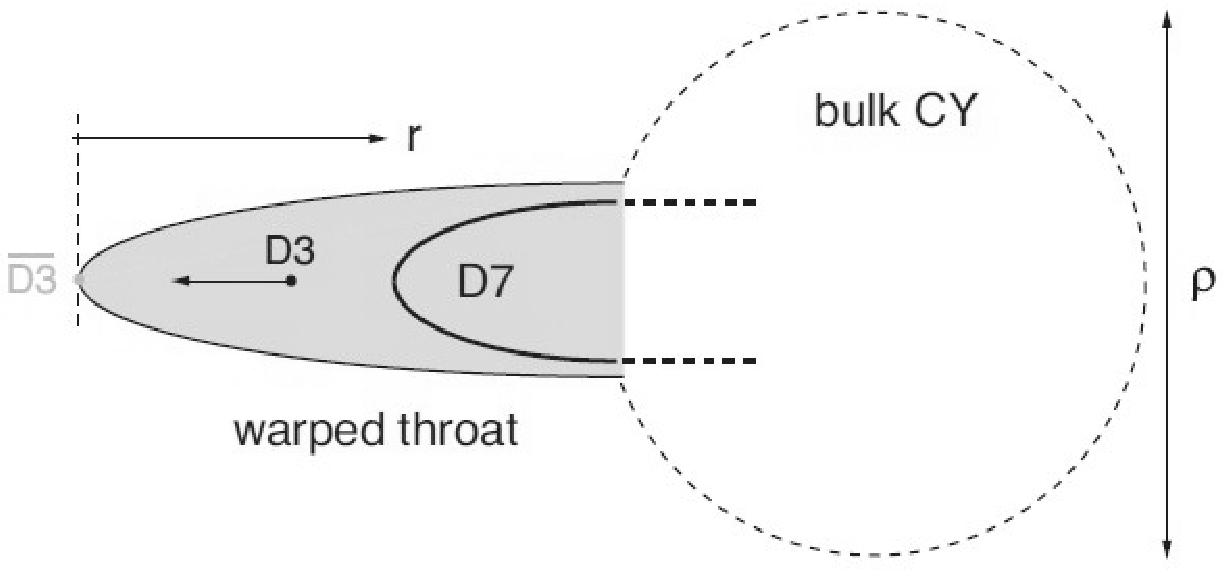,width=0.75\textwidth}{
A cartoon (taken from ref.~\cite{BDKMMM}) of the
configuration of D3 and D7 branes used in the superpotential
calculation. \label{d7 fig}}

To follow the explicit D3 dynamics we use the KKLT superpotential,
supplemented by the $w_i$-dependent corrections discussed above,
which are computed in ref. \cite{BDKMMM} for the case where
gaugino condensation occurs on a stack of D7's which wrap a cycle
extending into the throat, specified by the supersymmetric
embedding
\beq
    \prod_{i=1}^4 w_i^{p_i} = \mu^P \,,
\eeq
as in Fig.~(\ref{d7 fig}). The resulting superpotential can be
taken from \pref{WForm} and rewritten as
\beq \label{superp}
    W = W_0 + A(w_i) e^{-a\rho}
\eeq
where
\beq
    A(w_i) = A_0\left( 1 - {\prod_{i=1}^4 w_i^{p_i}\over
    \mu^P}\right)^{1/N_{D7}}
\label{sp}
\eeq
with $P=\sum_i p_i$ and $N_{D7}$ the number of D7 branes. In the
special case of $N_{D7}=1$ there is no gaugino condensation, but
the Euclidean D3-brane mechanism \cite{Witten} could instead
apply. This superpotential can be thought of as a function of the
distance between the D3 brane and the 4-cycle.  In principle,
there is a connection between the value of $a$ and the volume of
the 4-cycle \cite{BDKMMM}, $a=2 T_3 V_{\Sigma_4}/N_{D7}$, but for
the pure conifold this volume diverges, and would only be cut off
by the gluing of the throat to a bulk Calabi-Yau. Since this
procedure is model-dependent, we are free to consider $a$ as a
free parameter.

\subsection{D3 dynamics}

To compute the resulting D3 dynamics we first need to compute the
F-term potential, given by the usual supergravity formula
\beq \label{VFForm}
    V_F = e^{\kappa_4^2 K}\left( K^{\bar b a}D_a W\overline{D_b W}
    -3\kappa_4^2|W|^2\right)
\eeq
where the indices $a,b$ run over the complex fields $\rho,w_i$. It
is possible to show that the inverse K\"ahler metric, $K^{a\bar
b}$, is simple when expressed in block form, where the $\rho$ and
$w_i$ fields are written in separate blocks. Defining the volume
modulus as $R = \rho+\bar\rho - c k(w_i,\bar w_i)$, we find that
$K^{a\bar b}$ has the form
\beq
    K^{\bar b a} = {R\over 3}\left(\begin{array}{c|c} R +
    ck_{,\bar \imath}k^{\bar \imath j}k_{,j} &
    k_{,\bar \imath}k^{\bar \imath a}\cr
    \hline
    k^{\bar b i}k_{, i} & \frac{1}{c} k^{\bar b a}\cr
    \end{array}\right)
\label{Kinv}
\eeq
where $k^{\bar \imath j}$ is the inverse of $k_{i\bar \jmath} =
\partial_i \partial_{\overline\jmath} k$. Furthermore, using the K\"ahler
potential (\ref{kpcy}) we find that
\beq
    k_{,\bar \imath}k^{\bar \imath a} = \frac32 w^a;\qquad
    k^{\bar b i}k_{, i} = \frac32 \bar w^b; \qquad
    k_{,\bar \imath}k^{\bar \imath j}k_{,j} = r^2 \,,
\eeq
implying that the combination $R + c\, k_{,\bar \imath}\, k^{\bar
\imath j}k_{,j}$ appearing in (\ref{Kinv}) is simply $\rho+\bar\rho$.

Using these simplifications, the F-term potential takes the form
\beqa \label{W}
    V_F &=& {\kappa_4^2\over 3 R^2}\left[\phantom{\frac13}\!\!\!\!
    (\rho+\bar\rho)|W_{,\rho}|^2\right. -3(\overline W W_{,\rho} + {\rm c.c.} )
        \\
    &&\qquad\qquad\qquad +\left.\frac32
    \left(\overline W_{,\bar\rho} w^j W_{,j} + {\rm c.c.}\right)
    +\frac{1}{c} k^{\bar i j} \overline W_{,\bar i} W_{,j}\right] \nn \\
    &=& {\kappa_4^2\over 3 R^2}\Biggl[ \left[(\rho+\bar\rho)a^2+6a\right]
    |A|^2e^{-2a(\rho+\bar\rho)} + 3aW_0(Ae^{-a\rho} + \bar Ae^{-a\bar\rho})
    \\
    &&\qquad\qquad\qquad
    -\frac32ae^{-a(\rho+\bar\rho)}
        \left(\bar A w^j A_{,j} + {\rm c.c.}\right)
        + \frac{1}{c}  k^{\bar \imath j}\overline A_{,\bar \imath} A_{,j}
    e^{-a(\rho+\bar\rho)}\Biggr] \,, \nn
\eeqa
which vanishes for a purely constant superpotential ($a = A_i = 0$).
The first line of (\ref{W}) can be recognized as the KKLT
potential before doing any uplifting. The second line in (\ref{W})
contains the new contributions due to the $w_i$-dependent
superpotential corrections. The KKLT potential would give an AdS
minimum at $\sigma_{0}$ where $DW=0$ and
\beq \label{KKLTvac}
    W_0 = -A_0e^{-a\sigma_{0}}  \left(1+\frac23
    a\sigma_{0} \right), \quad \quad \quad V_{AdS} = -\frac{ a^2 A_0^2
    e^{-2a \sigma_{0}}}{6 \sigma_{0}} \,.
\eeq

\subsubsection*{The Ouyang embedding}

We now focus on the simplest embedding, discussed by Ouyang
\cite{Ouyang}, in which $p_1=1$ and $p_i=0$ for $i>1$. For
simplicity we also take $N_{D7}=1$.  We show that the
superpotential corrections to $V_F$,  denoted by $\delta V_F$, by
themselves do not uplift, because they vanish when the polar
angles of the $T^{1,1}$ space take their energetically preferred
values, $\theta_i=0$.  At small $\theta_i$, the $F$-term
contribution takes the form\footnote{We give an explicit formula
for the full $F$-term potential in the next section.}
\beq
    \delta V_F = M_{11} (\theta_1^2 + \theta_2^2)
    + M_{12}\cos\left(\sfrac12 \tilde\psi\right) \theta_1\theta_2 +
    \dots
\eeq
where $\tilde\psi = \psi-\phi_1-\phi_2$. The determinant of the
$\theta_i$ mass matrix is
\beq
    \det M_{\theta\theta} =  M_{11}^2 -
    \sfrac14\cos^2 \left( \sfrac12\tilde\psi \right)
    M_{12}^2
    \propto 1 - a^2 c^2 \mu^2 r
\label{Meq}
\eeq
when evaluated at the supersymmetric KKLT minimum, eq.\
(\ref{KKLTvac}), and when we take $\cos ( \sfrac12\tilde\psi ) =
\pm 1$, which minimizes the potential for $\tilde\psi$.

The following argument shows that for reasonable values of $g_s$,
$a^2 c^2 \mu^2 r_0 \ll 1$ (when $r$ is evaluated at the bottom of
the throat), and hence $\theta_i=0$ indeed minimizes the F-term
correction to the potential at a value where it vanishes. Suppose
$R_{CY}$ is the Calabi-Yau radius, and $r_0 = \xi_0 L \sim \xi_0
g_s^{1/4}/M_s$ in terms of the AdS curvature scale $L$ and the
warp factor at the bottom of the throat, $\xi_0 = h_0^{-1/4}$.
Using $c \sim T_3/M_p^2\sim 1/(g_s R_{CY}^6 M_s^4)$, $a = 2\pi/N$
for gaugino condensation of an $SU(N)$ gauge theory, and $\mu^2 <
R_{CY}^3$, we obtain
\beq
    a^2 c^2 \mu^2 r_0 < \left( \frac{2\, \pi}{N} \right)^2 { \xi_0\over
    g_s^{7/4} (R_{CY}\, M_s)^9} \label{estimate}
\eeq
We need $R_{CY} M_s \gg 1$ for the validity of the low-energy
effective theory, and since strong warping implies $\xi_0\ll 1$,
the right-hand-side of (\ref{estimate}) is generically $\ll 1$,
unless the string coupling is taken to be much smaller than its
normally assumed range of values. This shows that the curvature of
the potential in the $\theta_i$ directions is positive at
$\theta_i=0$, and implies that $\delta V_F$ has a local minimum at
the poles of the $S^3$ within the $T^{1,1}$, along which $\delta
V_F$ vanishes, for any values of the other coordinates within the
conifold. Thus, the D3 brane likes to move to these poles along
which the D3-dependence of the superpotential has no effect on the
energy of the itinerant D3 brane.

We can repeat the previous argument for larger values of $r$ to
see whether the situation can change higher in the throat. But
even assuming that $r\sim\mu^{2/3}\sim R_{CY}$, the bound
(\ref{estimate}) is only softened to
\beq
    a^2 c^2 \mu^2 r < \left(\frac{2\, \pi}{N} \right)^2
    {  1\over g_s^{2} (R_{CY}\, M_s)^8}
\label{estimate2}
\eeq
Even without the warp factor on the right hand side, it would seem
that having a large enough value of $R_{CY}\, M_s$ to justify the
effective field theory approach makes it impossible to violate
(\ref{estimate2}) without taking $g_s \ll (Ms R_{CY})^{-4}$ ---
much smaller than the ${\cal O}(0.1)$ values usually entertained.

\section{Dilaton corrections and uplifting}
\label{sec: uplifting}

Interestingly there is a competing effect which could cause the
brane to stabilize at nonvanishing polar angles, giving a positive
uplifting energy to the brane.
To study this we need to carefully analyse the
backreactions of the D7 brane on the geometry. These backreactions will
lead to possible running of the dilaton that will help us to study susy
breaking effects more precisely.

\subsection{D7 brane dynamics, running dilaton and D3 potential}

The internal metric
\pref{fimet} alongwith the warp factor $h$ captures the effect of the
D7 brane on the background volume $e^{4u}$, but the D7 brane also
distorts the axion-dilaton background. This distortion is
straightforward to work out using Ouyang's embedding \cite{Ouyang}.
Denoting the axion-dilaton by $\tau$ we see that both $\tau$ and the
axion $\tilde \phi$ can be denoted in the following way:
\beq \label{axtau}
\tau ~ = ~ {i\over g_s} ~ +{N_{D7}\over 2\pi i}~{\rm log}~w_1, ~~~~~~ \tilde\phi ~ = ~ {N_{D7}\over 4\pi}~(\psi -
\phi_1 - \phi_2)
\eeq
where $w_1$ is defined in \pref{wdef} and $N_{D7}$ is the number of D7 branes.
{}From \pref{axtau} it is easy to see that the dilaton for our case is given
by\footnote{For more general D7-brane configurations, the exact form of the dilaton
profile may be different, but the sign of the logarithmic dependence is expected to be robust,
since it encodes the information that adding flavors makes the the gauge
theory less asymptotically free.  We thank P.\ Ouyang for this
observation.}
\beq \label{dialton}
        e^{-\Phi} = {1\over g_s} - {N_{D7}\over 2\pi}
    \log\left({r^{3/2}\over\mu}\sin{\theta_1\over 2}\,
    \sin{\theta_1\over 2}\right) \,.
\eeq
which is exactly the same as the one derived in \cite{Ouyang}.\footnote{Ref.\ \cite{Ouyang} omits the factor $1/\mu$
in the argument of the $\log$, but it is clear that it is the
location of the D3 brane relative to that of the 4-cycle where the
D7 is wrapped which is relevant.}\ \
This is not surprising because we used non-trivial D7 monodromies to derive the $\tau$ background.
However this cannot be the complete story because of the SUSY-breaking effects that we discussed in sec. 2.
The two SUSY-breaking effects---existence of non-primitive fluxes
and D7 branes---rely explicitly on the fact
that the sizes of the two-spheres (parametrised by ($\theta_1, \phi_1$) and ($\theta_2, \phi_2$)) are unequal.
Therefore this should modify the dilaton behavior \pref{dialton} in
such a way as to reflect these changes.

To quantify the changes, let us assume that radii of the two-spheres
are related by
\beq \label{radius}
F_1 ~- ~F_2 ~= ~ \epsilon f
\eeq
in \pref{fimet}
with $\epsilon \to 0$ being a small quantity and $f$  a function which may depend on all coordinates. For such
a small change, the dilaton behavior cannot be too different from the
one given above \pref{dialton}.
Also since we are almost in the conifold type regime of our geometry \pref{fimet}, the monodromy property of
the embedding D7 has to be given by ${\rm log}~w_1$. The simplest way to keep the monodromy effect of D7 intact is
to perturb the behavior of $N_{D7}$, as we are changing the sizes of the cycles from the conifold case keeping the
axion flux quanta same. Therefore the {\it integrated} flux over a
cycle, which measures $N_{D7}$, will change slightly.
Thus our first ansatz would be to have
\beq \label{seven}
N_{D7} ~ \to ~ N_{D7} ~ + ~ \delta N(\epsilon)
\eeq
where $\epsilon$ is the change in \pref{radius}.\footnote{Physically the above formula says that a change in the volume
of the cycles of a conifold geometry to go to the background \pref{fimet} is {\it equivalent} to the
scenario where we have {remained} in the
conifold set-up but effectively changed the number of seven-branes.
Clearly this will only work when we are
close to the conifold geometry as specified by \pref{radius}. However
one immediate advantage of \pref{seven} is that we
can exploit all the useful properties of the conifold set-up to analyse the
system, yet provide solutions for the background
\pref{fimet} in the limit \pref{radius}.}

A little thought  tells us that this still cannot be the full story even if $\epsilon$ is very small. To
see a possible contradiction, let us consider the following scenario. Imagine our background \pref{fimet}
comes from a full F-theory set-up. Then there would be multiple seven-branes (not all local with respect to
each other). In such a scenario, there always exists a point in the moduli space where the string coupling
can be made locally constant \cite{DM}.
In such a case the global as well as local monodromies all vanish making
\beq
e^{-\Phi} ~ = ~ {1\over g_s}
\eeq
with no running dilaton and $\delta N = 0$ in \pref{seven}. This background is similar to the background
of \cite{Dymarsky} because of locally-cancelled seven-brane effects. However \cite{Dymarsky} still has a running
dilaton because of unequal radii of the two-cycles \pref{radius}. Thus the change \pref{seven} cannot fully
account for the change in dilaton behavior even for $\epsilon \to 0$. We need something more.

It turns out that
the additional corrections to \pref{dialton} can  be derived from
ref.\ \cite{Dymarsky}.
In the absence of D7 branes,
Dymarsky {\it et al.} claim that the dilaton runs as
\beq \label{dilrun}
e^{-\Phi_1} ~ = ~ e^{\epsilon^2 f^2 I(r)} ~ + ~ {\cal O}(\epsilon^4)
\eeq
where $I(r)$ is a function that is given in \cite{Minasian}. We see that when $\epsilon = 0$ then $e^{-\Phi_1} =1$
in \pref{dilrun}, $\delta N = 0$ and the only ``running'' will be from the monodromy analysis
\pref{dialton}. Thus the actual running of the dilaton for our case will be given by $e^{-\Phi} + \delta e^{-\Phi}$
where
\beq \label{actdil}
\delta e^{-\Phi} ~ = ~ -{\delta N(\epsilon) \over 2\pi}~ \log\left({r^{3/2}\over\mu}\sin{\theta_1\over 2}\,
    \sin{\theta_1\over 2}\right) ~ + ~ \left(e^{\epsilon^2 f^2 I(r)}~-~1\right) ~ + ~ {\cal O}(\epsilon^4).
\eeq
We remind the reader that this behavior for the dilaton is strictly
valid only in the limit $\epsilon \to 0$ where
conifold ans\"atze could be used. For a finite difference in radius \pref{radius} the monodromy behavior of the
D7 brane is more involved, and a simple ansatz like \pref{seven} has to be corrected with additional terms.

Now using the fact that supersymmetry is spontaneously broken in our background
\pref{metric} with ${\tilde{g}}_{mn}$ given by \pref{fimet}, a D3 brane along
spacetime directions $x^{0,1,2,3}$ should see a nonzero potential from the
Dirac-Born-Infeld (DBI) and Chern-Simons (CS) part of its action.
The potential in string frame is given in terms of the warp factor $h$ and the running dilaton by
\beq \label{dpoten}
\delta V_O ~ = ~ T_3\left[\sqrt{g}(e^{-\Phi}+ \delta e^{-\Phi})~-~C_{0123}\right] ~ = ~ T_3~ h^{-1} \delta e^{-\Phi}
\eeq
where $\sqrt{g}$ is the determinant of the metric along $x^{0,1,2,3}$ directions, $C_{0123}=g_s^{-1} h^{-1}$ is
the fourform background when the D3 brane is a probe, and $T_3$ is the tension of the
D3 brane. We see that the potential would vanish when the sizes of the
two-spheres are equal, as one might
expect.

We can determine the potential contribution if we know the warp factor $h(r)$ for our case. It is
clear that the leading term of the warp factor will be given by
\beq \label{harmoni}
h(r) ~=~ {L^4\over r^4} + \delta h(r,\theta_i, \epsilon)
\eeq
where $\delta h$ is the additional subleading contributions that may depend on $\theta_i$ coordinates and the
difference of the two radii $F_1-F_2$.
Providing this contribution does not cancel that of other bulk
fields, it leads to the following contribution to the D3 potential in Einstein frame:
\beq
    \delta V_O = - {\delta N(\epsilon)\over 2\pi}\,{T_3 \xi_0^4\over R^2}\,
    \left(r\over r_0\right)^4\,
    \log\left({r^{3/2}\over\mu}\sin{\theta_1\over 2}\sin{\theta_1\over 2}\right) ~ + ~ {\cal O}(\epsilon^2)
\label{ouyang1}
\eeq
which must be added to the F-term potential. The existence of this
potential is a clear manifestation that SUSY is broken in our case
even after we switch off the gaugino condensate term. It
is also clear that the new contribution is minimized at
$\theta_1=\theta_2=\pi$, so now there will be competition between
$\delta V_F$ and $\delta V_O$, resulting in a minimum at some
nontrivial value of $\theta_1=\theta_2=\theta$.

To finish this section, we need to determine the
value of $f$ appearing in the radius formula \pref{radius} and the sign of $\delta N(\epsilon)$. For $f$, we see that
in the analysis of Dymarsky {\it et al.}~\cite{Dymarsky} supersymmetry is already broken
at the level of D3 brane without any extra D7 brane. In their analysis
\beq
F_1(r)~-~F_2(r) ~\propto ~{\cal U}
\eeq
and so we might expect $f$ in \pref{radius} to be
be related to ${\cal U}$. The harmonic function $h$ in \cite{Dymarsky}
is of the form $h = h_{KS} + {\cal O}({\cal U}^2)$ where
$h_{KS}$ is the corresponding harmonic function for the
Klebanov-Strassler model. This is  also consistent with our choice
of harmonic function.

For the sign of $\delta N(\epsilon)$ we can go to the limit of our geometry \pref{fimet} where the tip is given
by a resolved conifold with the resolution parameter being $\epsilon$ appearing in \pref{radius}. For this case
we expect $\delta N > 0$ and so $\delta V_O$ computed above is
positive, and is suitable for uplifting.

\subsection{Minimization and uplifting}

We can explicitly integrate out the angular degrees of freedom
using the fact that $\delta V_F$ and $\delta V_0$ take the form
\beqa \label{Vieq}
    \delta V_F &=& V_1\sin^2\sfrac12\theta + V_2
    \sin^4\sfrac12\theta\\
    \delta V_O &=& V_O \log\left({r^{3/2}\over \mu}
    \sin^2\sfrac12\theta\right) ~ + ~ {\cal O}(\epsilon^2)
\eeqa
where
\beqa \label{ouyang}
    V_O &=& - {\delta N(\epsilon)\over 2\pi}\,{T_3 \, \xi_0^4\over
    R^2}\left(r\over r_0\right)^4\\
    V_1 &=& {\kappa_4^2\, A_0^2\, r\, e^{-2 a\sigma}\over 3\, \mu^2\, c \,R^2}
    \left( 3 - a \, \mu\,  c\, \sqrt{r}\cos\sfrac12\tilde\psi\left( 9 + 4\,a\,\sigma
    + 6{W_0\over A} e^{a\sigma}\right)\right)\\
    V_2 &=& {\kappa_4^2 A_0^2 \,  r \, e^{-2 a\sigma}\over 12\, \mu^2 \,c\, R^2}
    \left(-3 + a\, c\, r^2\,(12 + 8 \,a\,\sigma)\right)
\label{V12}
\eeqa
Minimizing $\delta V_F + \delta V_0$ over $\theta$ determines the
nontrivial angular minimum
\beq
    \sin^2 \left( \sfrac12\theta \right)
    = \min\left( {-V_1 + \sqrt{V_1^2-8 V_2 V_O}
    \over 4 V_2},\ 1\right)
\label{angle}
\eeq
This is the extra contribution
which will uplift the KKLT
potential to a nonnegative vacuum energy at its
minimum.\footnote{Using the KKLT minimization condition $W_0 =
-A_0e^{-a\sigma}(1+\frac23 a\sigma)$ to simplify the coefficient
of $\cos\sfrac12\tilde\psi$, we again see that the potential is
minimized when $\tilde\psi=0$, as we also showed after eq.\
(\ref{Meq})}

The correction $\delta V_F + \delta V_O$, evaluated at
(\ref{angle}), must be added to the KKLT potential
\beq
    V_{\sss KKLT} = {2\, \kappa_4^2\,  a \, A_0^2e^{-2a\sigma}\over R^2}\left(
    1 + \sfrac13 a\sigma  + {W_0\over A_0}e^{a\sigma}\right)
\label{KKLT}
\eeq
to obtain the full perturbed potential for the D3 brane and the
K\"ahler modulus.  We will see that the brane experiences a force
pushing it to the bottom of the throat, which we assume to be at
$r=r_0$.\footnote{A more accurate treatment would be to redo the
above calculations for the deformed Klebanov-Strassler throat,
or more generally for the resolved deformed case (\ref{fimet})
using the full running dilaton behavior, but
cutting off the throat at $r=r_0$.  We have
done these calculations for the deformed throat using the running 
dilaton ans\"atze \pref{actdil}, 
but did not
see any interesting differences relative to this simpler
treatment.} To study the problem of stabilizing $\sigma$ and
uplifting the potential at the minimum to nonnegative values, we
now restrict ourselves to the potential at $r=r_0$.

It is easy to see that the addition of $\delta V_O$ can be used to
raise the minimum of the potential to positive or zero values, by
comparing to the potential which arises in the KKLT procedure of
adding a $\overline{\rm D3}$ brane. The effect of a
$\overline{\rm D3}$ is to add a term
\beq
    \delta V_{\overline{\rm D3}} = {2 \xi_0^4 T_3\over R^2}
\label{antibrane}
\eeq
to the unlifted potential $V_{\sss KKLT}$.  The correction
(\ref{ouyang}) from the D3, has the same form, except for some
additional mild $\sigma$-dependence coming from the logarithm,
once (\ref{angle}) is imposed.  In fig.\ \ref{pots_fig} we show
the full potential as a function of the K\"ahler modulus for the
parameters (in $M_P$ units) $A_0=1,\ a=0.1,\ W_0=-10^{-4},\ r_0 =
0.1,\ c= 10^{-4},\ \mu=10$, and the warped brane tension
alongwith $\delta N(\epsilon)$ is tuned
to $\delta N(\epsilon) \xi_0^4 T_3 = 7.3\times 10^{-11}$ to get a Minkowski minimum
at $\sigma_0 = 115.8$.

\EPSFIGURE[ht]{Vsigma.eps,width=0.75\textwidth}
{The uplifted potential as a function of K\"ahler modulus,
for $r=r_0$ (bottom of the throat).
\label{pots_fig}}

Having demonstrated that the minimum can be uplifted to a positive
energy, we can now consider fluctuations of the brane from the
bottom of the throat and show that it is indeed driven to $r=r_0$.
We show the potential as a function of $r$ when $\sigma$ is at its
minimum-energy value, in figure \ref{potr_fig}.  The fact that the
potential turns over and goes to zero at some value of $r$ is
understandable, because of the form of the superpotential
(\ref{sp}), which vanishes at some maximum value of $r$, for fixed
angles.  We have checked that this is  exponentially close (of
order  $e^{-a\sigma}$) to the value of $r$ at which the potential
shown in figure \ref{potr_fig} vanishes. Without the competition
between the superpotential and the correction
(\ref{ouyang1}), we do not get any uplifting effect.

\EPSFIGURE[ht]{Vr.eps,width=0.75\textwidth}
{The potential as a function of $r$ at the stationary
value of $\sigma$.\label{potr_fig}}

\subsubsection*{Potential near the bottom of the throat}

For completeness, we note that the form of the full potential for the
brane simplifies in the region $r\ll \mu^{2/3}$. In this region, $V_O
\sim r^4$ while $V_i\sim r$ in (\ref{ouyang}-\ref{V12}), and we can
approximate (\ref{angle}) by Taylor-expanding in $V_0$, giving
\beq
    \sin^2 \left( \sfrac12\theta \right) = -{V_O\over V_1} = {|V_O|\over V_1}
\eeq
This results in the potential
\beqa
    V(r) &=& |V_O|\left[ 1 - {|V_O|\over V_1}  \,
    \log\left( {r^{3/2}\over \mu} \right) \right]
    + V_{\sss KKLT}
    \label{newpot}
\eeqa
where $V_{\sss KKLT}$ depends on $r$ only weakly, $V_{\sss KKLT} =
-C/R^2$, since $R = (2\sigma - cr^2)$ and $C$ is a positive
constant.  Of course this ``weak'' dependence was the origin of
the severe $\eta$ problem of the KKLMMT model, but here the $r^2$
dependence is removed, because the first term in (\ref{newpot}) is
tuned to go like $C/R^2$ as $r\to r_0$, so that the potential is
zero at its minimum.  This causes the $r^2$ dependence in $R$ to
be subleading to the main $r^4$ behavior of the potential.

\subsection{Slow roll brane-antibrane inflation? }

We next explore in a preliminary way whether the potential
corrections considered above can be used to obtain slow roll
inflation. First we consider the evolution for small $r$, where we
have seen that the potential varies for small $r$ like $r^4$.
However, the kinetic term for the inflaton is in this case given
by \cite{KKLMMT, realistic}
\beq
    M_p^2{6\, c\,\sigma\over R^2}\,\dot r^2 \sim {T_3\over \sigma}\dot
    r^2\,.
\eeq
In the limit where $cr^2\ll 2\sigma$ and $R\cong 2\sigma$, this
implies that the canonically normalized field, $\psi$, satisfies
\beq
    \frac{\psi}{M_p} = \sqrt{3\,c\, r^2\over\sigma} < \sqrt{6} \,.
\label{maxfield}
\eeq
The inequality in (\ref{maxfield}) follows because $R$ can never
become negative (being related to the physical size of the
Calabi-Yau), and so $r$ cannot exceed $\sqrt{2\sigma/c}$.

On the other hand, in order to obtain chaotic inflation with a
$\psi^4$ potential, one needs to ensure that the $\eta$ parameter,
$M_p^2 V''/V = 12 M_p^2/\psi^2$, is much smaller than unity, with
inflation ending when $\eta\sim 1$ --- that is, when $\psi =
\sqrt{12} M_p$. This shows that (\ref{maxfield}) is incompatible
with chaotic slow-roll inflation: the field is already rolling too
fast even for the largest values of $r$.

Next, we consider whether it is possible to inflate from the top
of the potential where it has a local maximum.  This is equivalent
to asking whether we can use the negative curvature of our new
contribution to the potential to cancel the positive curvature
which comes from expanding an antibrane contribution to the
potential (\ref{antibrane}) in $r$.  In the approximation of
ignoring the brane-antibrane Coulombic attraction term, there is
no difference between (\ref{antibrane}) and the KKLT potential
(\ref{KKLT}) as far as their $r$-dependence is concerned.

At the local maximum of the potential shown in figure
\ref{potr_fig}, the value of $\eta$ is
\beq
    |\eta| \sim {\sigma\over c} \left|{V_{,rr}\over V}\right|
    \sim {\sigma\over c\, r_m^2} \gg 1
\eeq
where $r_m\sim 1$ is the value of $r$ at the maximum. To obtain a
smaller value of $\eta$, one should choose parameters such that
$r_m$ becomes larger.  However, the condition $R>0$ ensures that
$r$ must satisfy the constraint $r<\sqrt{2\sigma/c}$, and this
shows that in the best case $\eta$ can at most be of order $1$,
whereas we need $\eta\sim 1/N_e$ with $N_e\sim 60$ being the
number of $e$-foldings of inflation.

It is possible that slow-roll inflation might become possible in
more complicated constructions, such as if supersymmetry-breaking
fluxes are turned on on the D7 branes. In this case it may be
possible to balance the resulting $D$-term potential, which acts
to attract the D3 brane towards the D7, with the $F$-term
potential considered here. We leave a more detailed study of the
interplay of these D3 forces to future work.

\section{Applications to orbifolds $M \times T^2/\Gamma$}
\label{sec: dterm}

We next specialize to toroidal geometries for which the internal
geometry is locally a product of the 4-manifold, $M$, and an
orbifolded 2-torus, $T^2/{\Gamma}$, with the various D7's wrapping
$M$ and located at the fixed points, $z = z_s$, of the orbifold.
(In this section we denote the complex coordinates on $M$ by $y^i$
and those on $T^2/\Gamma$ by $z$, so that the coordinates on $M
\times T^2/\Gamma$ are $\{u^m \} = \{ y^i,z \}$.) For instance for
the case of $K3 \times T^2/\mathbb{Z}_2$ we would have 4 D7's and
an orientifold plane -- each of which wraps $K3$ -- located at
each of 4 fixed points on $T^2/\mathbb{Z}_2$. Our interest in this
instance is in the motion of an itinerant D3 brane within the flat
toroidal dimensions.

\subsection{The gauge coupling function}

The dependence of the perturbed warp factor, $h(u)$, on the D3
position, $z$, is found by solving the perturbed supergravity
equations in the bulk. Taking $h \to h_0 + \delta h$, we have
\be \label{nabladeltaheqn}
    \tilde\nabla_u^2 \delta h(u) = -2\kappa_{10}^2 T_3
    \left[ \frac{\delta^6 (u-z)}{\sqrt{\tilde g}_6} - \sigma_b \right] \,,
\ee
where $\sigma_b$ is the background charge density coming from the
adjustments made by all of the other sources in response to the
presence of the D3 in order to maintain the topological condition
that the integration over the left-hand-side vanish for a compact
space. The authors of ref.~\cite{BDKMMM} argue that the presence
of the $\sigma_b$ term implies the existence of a nonholomorphic
contribution to $\delta h$
\be \label{deltahingeneral}
    e^{4u} \, \delta h = G(u^m;z) +
    \overline G(\overline u^{\overline{m}};\overline z)
    + \frac{\kappa_{4}^2 T_3}{3} \, k(z,\overline z) \,,
\ee
with the last term precisely cancelling the nonholomorphic
contribution of $\delta e^{4u}$ in its contributions to the gauge
coupling function. (Here $e^u$ represents the breathing mode,
$\tilde{g}_{m\overline n} = e^{2u} g_{m\overline n}$.) The
resulting gauge coupling function is then determined by the
holomorphic contribution, $G(u^m,z)$, of the appropriate Green's
function, suitably integrated over the cycle, $\Sigma$, wrapped by
the corresponding D7 brane.

With these results the gauge coupling function for the D7 brane
located at $z = z_s$ on $T^2/\Gamma$ becomes $f_s(\rho,z) = \rho +
F_a(z)$, with the D3 position-dependence being
\bea \label{F vs z}
    F_s(z) &\propto& \cT_7 \,
    \int_\Sigma d^4y \sqrt{g_4} \;
    \delta\Bigl( e^{4u} \, h \Bigr)_{hol} \nn\\
    &=& \cT_7  \int_\Sigma d^4y \sqrt{g_4} \;
    G(y,z_s;z)  \,.
\eea
The integration of the 6D Green's function, $G(y,z)$, over the
volume of the 4-cycle $\Sigma$ simply converts the result into the
appropriate 2D Green's function. That is, integrating
eq.~\pref{nabladeltaheqn} over the 4-cycle and using
$\tilde\nabla^2_u = \tilde\nabla^2_y + \tilde\nabla^2_z$ implies
$\langle \delta h \rangle_\Sigma$ satisfies
\be
    \tilde \nabla^2_z \langle \delta h \rangle_\Sigma = - \frac{2
    \kappa_{10}^2 T_3}{\tilde{V}_\Sigma} \left[
    \frac{\delta^2(y - z)}{\sqrt{\tilde g_2}} -  \sigma_2 \right] \,,
\ee
where we define $\langle \cdots \rangle_\Sigma =
\tilde{V}_\Sigma^{-1} \int d^4y \sqrt{\tilde g_4} (\cdots)$,
$\tilde{V}_\Sigma$ is the volume of the 4-cycle computed with the
metric $\tilde{g}_{mn}$, $\tilde\nabla^2_z$ denotes the 2D
Laplacian and $\sigma_2 = \int_\Sigma d^4 y \sqrt{\tilde g_4} \;
\sigma_b$. For instance, for the torus defined by the lattice $y
\simeq y + 1 \simeq y + \tau$, with complex modulus $\tau = \tau_1
+ i \tau_2$ and K\"ahler metric $\exd s^2_{\scriptscriptstyle T} =
\exd y \, \exd \overline y$, the volume of the 2-torus is $V_T =
\tau_2$ and $\sigma_2 = 1/\tau_2$. The result for $\langle \delta
h \rangle_\Sigma$ then becomes
\be \label{Torusdeltah}
    \langle \delta h(w,\overline w) \rangle_\Sigma = -
    2\kappa_{4}^2 T_3 \tau_2
    \left[ \frac{(w - \overline w)^2}{8\tau_2} +
    \frac{1}{4\pi} \ln \Bigl| \vartheta_1 \left( \pi w | \tau
    \right) \Bigr|^2 \right] \,,
\ee
where $w = z - z'$ and we use $\kappa_{10}^2 = \kappa_4^2 V_6 =
\kappa^2_4 V_\Sigma \tau_2$. Again the nonholomorphic part is
proportional to the 2D K\"ahler potential, $k(w,\overline w)$,
which ref.~\cite{BDKMMM} argues is cancelled by the
nonholomorphic contribution of the back-reaction to $e^{4u}$.

A similar result holds when the two dimensions transverse to the
D7 are orbifolded, obtained by summing eq.~\pref{Torusdeltah} over
the appropriate image points:
\be \label{Orbifolddeltah}
    \langle \delta h(w,\overline w) \rangle_\Sigma = -
    2\kappa_{4}^2 T_3 \tau_2 \sum_p
    \left[ \frac{(w_p - \overline w_p)^2}{8\tau_2} +
    \frac{1}{4\pi} \ln \Bigl| \vartheta_1 \left( \pi w_p | \tau
    \right) \Bigr|^2 \right] \,,
\ee
where $w_p = z - g_p(z')$ and $g_p(z')$ denotes the action on the
D3 brane position, $z'$, of the discrete group elements, $g_p$,
with the sum running over all of the elements of the group. For
instance, for the orbifold $T^2/\mathbb{Z}_2$ defined by
identifying points under reflection of a square torus about the
origin, we have $g_+(z) = z$ and $g_-(z) = -z$ and $\tau = i$, and
so $w_+ = w = z - z'$ while $w_- = z + z'$.

Using this in expression \pref{F vs z} gives the following
expression for the dependence of $F_s(z)$ on the D3 position
\be \label{Fs}
    F_s(z) = C
    \sum_p \ln \Bigl( \vartheta_1 \left[ \pi (z_s - z_p) | \tau
    \right]  \Bigr) \,,
\ee
where $z$ denotes the position of the itinerant D3 brane in the
orbifold, $z_p$ is its image under the orbifold group elements,
and $z_s$ is the position of the D7 brane of interest. $C$ denotes
a constant whose detailed form is not crucial in what follows,
which is proportional to the tension of the D3.

\subsection{$F$-term potential: bulk fluxes}

Using this in the gaugino-condensation superpotential
\be \label{WForm1}
    W = W_0 + \sum_s A_s \exp \Bigl[ -a_s (\rho + F_s(z)) \Bigr] \,.
\ee
gives the low-energy expression of the forces on the D3 due to the
physics of modulus stabilization. As we discussed before, the
constant $W_0$ appears from the $G_3$ given in \pref{twne}, and
therefore expresses the effects of supersymmetry-breaking amongst
the higher-dimensional fluxes which stabilize some of the moduli,
while the exponential term contains the influence of gaugino
condensation (or the like) on various D7 branes, and involves the
dependence on the D3 position due to the back-reaction of the D3
onto the relevant gauge coupling strengths. The quantities $A_s$
and $a_s$ (which are related to $A_0$ and $a$ respectively in
\pref{superp} and \pref{sp} for special choices of $s$) are
$z$-independent constants which are calculable given the details
of the underlying physics.

\subsubsection*{Periodicity properties}

For later purposes it is instructive at this point to record a
subtlety regarding the periodicity properties of the above
expressions under the shifts $z \to z + 1$ and $z \to z + \tau$
which define the underlying torus, restricting for convenience to
the case of later interest: the orbifold $T^2/\mathbb{Z}_2$, with
$\mathbb{Z}_2$ acting as $z \to -z$. Using the periodicity
properties of the Jacobi $\vartheta$-function listed in the
Appendix, it can be shown that the quantities
\bea \label{fstorus}
    f_s(\rho,z) &=& \rho + \alpha \sum_p \ln \Bigl( \vartheta_1 \left[ \pi(z_s -
    z)|\tau \right] \Bigr) \nn\\
    &=& \rho + \alpha \ln \Bigl( \vartheta_1 \left[ \pi(z_s - z)|\tau
    \right] \vartheta_1 \left[ \pi(z_s + z)|\tau \right] \Bigr)
    \\ \label{Xtorus}
    \hbox{and} \qquad X &=& \rho + \overline\rho - \frac{\beta \,
    (z-\overline z)^2}{\tau_2} \,,
\eea
are invariant under the transformations
\bea \label{Periodicity}
    &&z \to z + 1 \qquad \hbox{and} \qquad \rho \to \rho \nn\\
    &&z \to z + \tau \qquad \hbox{and} \qquad \rho \to \rho +
    2i\beta\,(2z + \tau) \,,
\eea
provided the real constants $\alpha$ and $\beta$ are related by
$\beta = \pi \alpha$.

This shows that the $F$-term potential built from the above
superpotential is appropriately periodic under shifts of the D3
position, but only if the volume modulus, $\rho$, also shifts
appropriately. This coupling of the shifting of $\rho$ and $z$ due
to the nonperturbative $F$-term potential shows that the D3
modulus, $z$, transforms nontrivially under the classical shift
symmetry, $\rho \to \rho + i \epsilon$, which is broken by
anomalies down to a discrete subgroup under which both $\rho$ and
$z$ shift. This bears out the observation \cite{SSeq} that
symmetries can require the KKLT superpotential to depend on fields
other than just $\rho$, although it is interesting that the the
fields which are relevant are in this case the position moduli of
the itinerant D3 rather than charged multiplets living on the
branes. This cancellation between shifts of $z$ and $\rho$ is also
noted in ref.~\cite{HKLVZ}.

A novel feature of this realization of the symmetry is that it
does not involve any fields beyond $\rho$ and the D3 position
modulus, $z$. In particular it does not involve any charged chiral
fields on the branes, such as is often assumed. In fact, in the
limit that the D3 approaches the relevant D7 brane the dependence
of $W$ on $z$ has a natural interpretation from the point of view
of the charged fields which become part of the effective 4D field
theory as the mass of the D3--D7 string states become light. This
is because when the D3 is sufficiently close to the D7, the mass
of these states is nonzero but small enough to be included into
the low-energy 4D theory, through a contribution to the
superpotential of the form
\be \label{Wmassterm}
    W_{\scriptscriptstyle M} = \frac12 \, \mu_{ij}(z)
    \varphi^i \varphi^j \,
\ee
where the dependence of $\mu_{ij}(z)$ on the D3 position, $z$, is
calculable and arises because of the necessity of stretching the
D3--D7 strings as the D3 position changes. Furthermore, the $U(1)$
invariance of eq.~\pref{Wmassterm} ensures that $\mu_{ij}(z)$
necessarily has the right transformation property to be combined
into an invariant gaugino-condensation superpotential in
combination with $e^{-a\rho}$ \cite{SI}, along the lines discussed
in refs.~\cite{SSeq}, corresponding to what would be obtained if
the charged fields were integrated out. This shows why it can be
possible to achieve invariance using only the fields $z$ and
$\rho$.

\subsection{$D$ term potential: brane fluxes}

A $D$-term potential, eqs.~\pref{VDterm} and \pref{DForm}, can
also be generated for these compactifications if
supersymmetry-breaking magnetic fluxes are turned on on some of
the D7 branes. This type of $D$-term potential arises in
particular if the D3 is brought close enough to the relevant D7
that the D3-D7 string states become light enough to introduce
chiral multiplets whose scalar fields carry the charge of the D7
gauge group.

If we use the form of the K\"ahler potential, $\kappa_4^2
\partial_\rho K = -3/X$, and the gauge kinetic function, $f_s =
\rho + F_s(z)$, suggested by eqs.~\pref{fstorus} and
\pref{Xtorus}, then $V_D$ becomes
\be \label{VDForm}
    V_D = \sum_s \frac{V_{0s}}{f_s + \overline f_s} \left[
    \frac{3\, v_s}{X}  + \cD_s(\varphi,\overline\varphi) \right]^2 \,,
\ee
where $X = \rho + \overline\rho - \beta (z-\overline z)^2/\tau_2$,
the constant $V_{0s}$ is inversely proportional to the tension of
the relevant brane, as well as to the volume of the 4-cycle which
it wraps. The parameter $v_s$ is proportional to the strength of
the flux whose presence on brane `$s$' breaks supersymmetry. The
scalar fields, $\varphi$, here denote any light scalars which are
charged under the relevant gauge group and appear in the effective
4D theory. These might include light D3--D7 string states if the
D3 brane is sufficiently close to the relevant D7, but would not
if the D3--D7 separation should be too great.

\subsection{Uplifting}

The potential \pref{VDForm} was proposed in ref.~\cite{BKQ} as
being a potential source of uplifting to flat or anti-de Sitter
space, instead of using the supersymmetry-breaking anti-D3 brane
used by KKLT. However, as noted in \cite{BKQ} the success of this
proposal is model-dependent inasmuch as it relies on the minimum
of the complete scalar potential being at a place where some of
the $D$-terms are nonzero. As has since been emphasized
\cite{SSeq}, this requires at least one of the $F$-terms to also
be nonzero at the relevant minimum.

Both of these features are explicitly manifest in the potential $V
= V_F + V_D$, if $V_F$ and $V_D$ are computed with gaugino
condensation and supersymmetric flux breaking occuring on a single
D7 brane. In this case if we take an itinerant D3 brane which is
far enough from the D7 then the only relevant light fields are
$\rho$ and $z$ because all of the charged D3--D7 states are too
massive to be included in the effective 4D theory (and so
$\cD_s(\varphi, \overline\varphi) = 0$). In the absence of a
Fayet-Iliopoulos term $V_D = 0$, and at face value $V_F$ as
computed above is minimized (and vanishes) as the D3 moves towards
the D7 on which gaugino condensation and flux breaking occurs,
because the gauge kinetic function, $f_s$, diverges
logarithmically as $z \to z_s$. However, in reality the effective
description must change before the D3 and D7 can reach one another
because of the breakdown of the approximations used (such as the
necessity to include the D3--D7 states which become light in this
limit).

A less trivial situation would arise if a brane configuration
could be devised for which there is a $U(1)$ gauge group in the
low-energy theory for which all of the charged chiral multiplets,
$\varphi$, have the same sign charge. In this case the low-energy
theory has a $U(1)$ anomaly, whose Green-Schwarz cancellation
implies the existence of a Fayet-Iliopoulos term \cite{FIanomaly},
as the following argument shows. In such a case the
anomaly-cancelling mechanism implies that the 4D Lagrangian
contains the term $v \int B \wedge F$, where $v$ is a constant
with dimensions of mass, $F = \exd A$ and $B$ is the appropriate
two-form potential which shifts under the action of the anomalous
$U(1)$ gauge transformation. Together with the $B$-field kinetic
terms, this dualizes to a Lagrangian of the form
\be \label{4DGSEqn}
    S_{4dgs} = - \int d^4x  \left[ \sqrt{-g} \; \frac12 (\partial_\mu a
    - vA_\mu) (\partial^\mu a - vA^\mu) + c'\, a\, F \wedge F \right] \,,
\ee
where $a$ is the Goldstone mode dual to $B_{\mu\nu}$ in four
dimensions, and $c'$ is an appropriate constant. $N=1$
supersymmetry then implies that the corresponding SUSY multiplets
$A$ and $\psi$ only enter the K\"ahler potential of the low-energy
theory through the combination $\psi + \overline\psi - vA$ within
the K\"ahler function, $K$, where $\psi$ is the complex scalar
whose imaginary part is $a$ and $A$ is the vector multiplet
containing $A_\mu$. This leads to an FI term having the form $\xi
= - v\partial_\psi K = 3v/X$. Furthermore, eq.~\pref{4DGSEqn}
requires the gauge kinetic term for $A_\mu$ must contain a term
linear in $\psi$.

Of particular interest for us is the case where the relevant
scalar $a$ is the imaginary part of the volume modulus, $\rho$,
since we know that this field generically does appear linearly in
the gauge kinetic functions in type IIB compactifications. Since
this field comes from the component $C_{\mu\nu mn}$ of the RR
4-form field, the 4D Green-Schwarz term can be regarded as the
low-energy expression of the underlying 7-brane Chern-Simons
coupling
\be
    S_{cs} \propto \int C \wedge F \wedge \cF \,,
\ee
where $F_{\mu\nu}$ is the 4D gauge field for the anomalous $U(1)$
and as above $\cF_{mn}$ is the background flux whose SUSY-breaking
presence the FI term represents.

In such a case the $D$-term potential contains contributions from
both the FI term as well as the contributions of the charged
scalars, $\cD_s(\varphi,\overline\varphi)$. However the relative
sign of these contributions to $D_s$ is dictated by supersymmetry
and anomaly cancellation, and is such that the minimization with
respect to $\varphi$ cannot cancel $\cD_s$ against the FI term,
leading to a minimum at $\cD_s = 0$, with a surviving nonzero FI
term available to play a role in uplifting. It would clearly be of
considerable interest to realize this picture within a {\it bona
fide} string construction.

\subsection{Beyond linear backreaction?}

The derivation leading to eqs.~\pref{WForm}, \pref{VFForm} and
\pref{VDForm} for the potentials $V_F$ and $V_D$, includes the
D3-brane position through its dependence on the gauge kinetic
function, $f_s = \rho + F_s$, and on the K\"ahler variable, $X$,
which themselves depend on the D3 position through
eqs.~\pref{deltahingeneral} and \pref{Kahlerform} (or
\pref{Xtorus}). This dependence arises due to the back-reaction on
the bulk fields of the D3 position, and was computed by
linearizing about the D3-independent background. This leads us to
ask whether the domain of validity of the 4D potential must also
be restricted to linear order in $F$ and $\delta X = X - \rho -
\overline\rho$.

Part of the virtue of having a formulation in terms of 4D
supergravity lies in the various nonrenormalization theorems which
such theories enjoy \cite{NonRenorm,NRTheoremsSym}. For
holomorphic quantities like the gauge kinetic function and
superpotential, these theorems often allow the extension of
nominally low-order results beyond the domain of their initial
derivation. In particular, since nonrenormalization theorems often
restrict the corrections to the gauge kinetic function to arise
only at lowest order, we expect that it may be a good
approximation to keep the full dependence of $f_s = \rho + F_s$ on
$z$, without having to linearize results to lowest order in $F$.

Similar arguments are more difficult to make for $X$, however,
since corrections to the K\"ahler function of the low-energy 4D
supergravity are typically not protected from receiving
perturbative corrections. Indeed, it can happen that interesting
and qualitatively new kinds of minima actually do arise for the
scalar potential once the leading such corrections are taken into
account \cite{BCPQ}.

\section{Conclusions}

Building on the work of \cite{BDKMMM}, we have shown that the KKLT
stabilization mechanism, with the addition of a mobile D3 brane,
necessarily involves extra superpotential and dilaton background
corrections. These are a consequence of the same mechanism that
stabilizes the K\"ahler modulus, either Euclidean D3 branes or
gaugino condensation. In a scenario like brane-antibrane inflation
where D3 branes are involved it is necessary to add these
corrections.

The new corrections depend upon which 4-cycle in the KS-throat is
wrapped by the D7 brane, or stack of D7 branes.  In the present
work we have focused on a particularly simple choice of 4-cycle,
and the case of a single D7 brane.  Our preliminary study of other
choices indicate that they have similar qualitative behavior to
the simple case we studied.

A major motivation for studying this system was to determine
whether the superpotential correction can be fine-tuned to
ameliorate the $\eta$ problem of brane-antibrane slow roll
inflation \`a la KKLMMT \cite{KKLMMT}.  A yet more fortunate
outcome would  have been to find that our potential supports slow
roll inflation by itself, even without an antibrane. We found that
neither of these possibilities could be realized.  Either one
would have required large values of $r$, inconsistent with the
requirement $cr^2\ll 2\sigma$, needed in order to keep the volume
of the extra dimensions sufficiently large that the low-energy
effective description can be trusted.  It has been suggested that
these problems can be overcome in a more elaborate related
background, the {\it full} resolved warped deformed conifold \cite{Dymarsky},
although it would be worth extending their analysis
to a case for which all moduli are stabilized.\\

Finally, we examine a simple toroidal example and exhibit the $F$
and $D$ term potentials which express in the low-energy 4D theory
various forces on a mobile D3 brane. We show how the D3 position
modulus can play the role of the field which ensures the
invariance of the superpotential under otherwise-puzzling
symmetries. We imagine the resulting potential could be useful for
exploring D3--D7 inflationary models in more detail.

\section*{Acknowledgments}

We thank D. Baumann, J. Conlon, A. Davis, A. Frey, R. Kallosh, L.
McAllister, P. Ouyang, M. Postma and F. Quevedo for helpful
conversations, as well as the Banff International Research Station
and the Benasque Center for Physics for providing us with the
extremely pleasant time and place which made this work possible.
All four authors acknowledge support from the Natural Sciences and
Engineering Research Council of Canada, and CB acknowledges
additional research support from McMaster University and the
Killam Foundation. Research at Perimeter Institute is supported in
part by the Government of Canada through NSERC and by the Province
of Ontario through MEDT.

\appendix

\section{Theta functions}

We record in this appendix some of the properties of the Jacobi
$\vartheta$-functions we use in the main text. We use the
definition
\bea
    \vartheta_1(z|\tau) &=& -i \sum_{n = -\infty}^\infty (-)^n
    q^{\left( n+\frac12 \right)^2} \, e^{i(2n+1)z} \nn\\
    &=& 2 \sum_{n = 0}^\infty (-)^n q^{ \left( n+\frac12\right)^2}
    \sin \Bigl[ (2n+1) z \Bigr]
    \,,
\eea
where $q = e^{i\pi\tau}$. This definition ensures that
$\vartheta_1(-z|\tau) = - \vartheta_1(z|\tau)$, as well as the
toroidal periodicity properties
\bea
    \vartheta_1(z + n\pi|\tau) &=& (-)^n \vartheta_1(z|\tau) \nn\\
    \hbox{and} \quad
    \vartheta_1(z - n\pi \tau| \tau) &=& (-)^n q^{-n^2} e^{2niz}
    \vartheta_1(z|\tau) \,,
\eea
for any integer $n$. Similarly, the identity
\be
    \frac{\vartheta'_1(z|\tau)}{\vartheta_1(z|\tau)} = \cot z + 4
    \sum_{n=1}^\infty \left( \frac{q^{2n}}{1 - q^{2n}} \right)
    \sin(2nz) \,,
\ee
where $\vartheta'_1(z|\tau) = \exd \vartheta_1(z|\tau)/\exd z$,
implies that near $z = 0$ we have $\vartheta_1(z|\tau) = z +
O(z^3)$.

\section{4D anomaly cancellation}

In this appendix we confirm the relative sign between the
contributions to the $D$-terms from the Fayet-Iliopoulos (FI) term
and from the charged matter multiplets. We show that in the
special case where all of the matter multiplets share the same
charge the resulting potential is minimized by having the charged
scalars vanish, leaving the D3 dynamics governed by the FI term,
as argued in ref.~\cite{BKQ}.

We start with the 4D $U(1)$ anomaly due to a collection of
fermions, all of which carry the same $U(1)$ charge $q$. By an
appropriate choice of counterterms the variation of the quantum
action under such an anomaly can be written in the form:
\be
    \delta S = \frac{Aq}{16\pi^2}
    \int \exd^4x \,\omega \epsilon^{\mu\nu\lambda\rho}
    F_{\mu\nu} F_{\lambda\rho} \,,
\ee
where $\omega$ is the $U(1)$ symmetry transformation parameters
and $A$ is a positive calculable constant.

Within the 4D Green-Schwarz mechanism this anomaly is cancelled by
the presence of a local interaction, given the presence of a
2-form gauge potential, $B_{\mu\nu}$, with an action
\be \label{GSaction1}
    S_{\scriptscriptstyle GS} = - \int \exd^4x \left[
    \frac{1}{12} \, \sqrt{-g} \, H^{\mu\nu\lambda}
    H_{\mu\nu\lambda} + k \epsilon^{\mu\nu\lambda\rho} B_{\mu\nu}
    F_{\lambda\rho} \right] \,,
\ee
where the field strength
\be \label{FieldStrength}
    H_{\mu\nu\lambda} = \partial_\mu B_{\nu\lambda} - \kappa A_\mu
    F_{\nu\lambda} + \hbox{cyclic} \,,
\ee
is invariant under the following $U(1)$ gauge transformations
\be
    \delta A_\mu = \partial_\mu \omega \qquad \hbox{and}
    \qquad \delta B_{\mu\nu} =  \omega \kappa F_{\mu\nu} \,,
\ee
and $\kappa$ is a positive dimensionful constant. This action
cancels the anomaly for an appropriate choice of $k$ because its
variation under the $U(1)$ transformation is
\be
    \delta S_{\scriptscriptstyle GS} = - \kappa k \int \exd^4x \;
    \omega \epsilon^{\mu\nu\lambda\rho} F_{\mu\nu} F_{\lambda\rho}
    \,.
\ee
This cancels the fermionic anomaly provided $\kappa k = Aq/(16
\pi^2)$.

The connection to supersymmetric $D$-terms is best seen once the
2-form field is dualized, leading in 4D to a scalar field, $a$.
This duality is most easily performed by rewriting the
Green-Schwarz action as
\bea \label{GSaction2}
    S_{\scriptscriptstyle GS} &=& - \int \exd^4x \left[
    \frac{1}{12} \, \sqrt{-g} \, H^{\mu\nu\lambda}
    H_{\mu\nu\lambda} + \frac{2k}{3} \epsilon^{\mu\nu\lambda\rho} A_{\mu}
    H_{\nu\lambda\rho} \right. \nn\\
    &&\qquad\qquad\qquad\qquad\qquad \left. + \frac{1}{6}\, a\,
     \epsilon^{\mu\nu\lambda\rho}
    \left( \partial_\mu H_{\nu\lambda\rho} + \frac{3\kappa}{2} F_{\mu\nu}
    F_{\lambda\rho} \right) \right] \,,
\eea
and regarding the functional integral to be over the fields
$H_{\mu\nu\lambda}$ and $a$, rather than $B_{\mu\nu}$. The
equivalence of this form with eq.~\pref{GSaction1} is seen by
performing the functional integral over the field $a$, which acts
as a Lagrange multiplier enforcing the Bianchi identity:
\be
    \epsilon^{\mu\nu\lambda\rho} \partial_\mu H_{\nu\lambda\rho} =
    - \frac{3\kappa}{2} \epsilon^{\mu\nu\lambda\rho} F_{\mu\nu}
    F_{\lambda\rho} \,.
\ee
This has eq.~\pref{FieldStrength} as its local solution, allowing
the functional integral over $H_{\mu\nu\lambda}$ to be traded for
an integral over $B_{\mu\nu}$, weighted by the action
\pref{GSaction1}.

The dual formulation is obtained by performing the functional
integrals in the opposite order, first integrating over
$H_{\mu\nu\lambda}$ to leave an action in terms of the scalar
field $a$. Since the integral over $H_{\mu\nu\lambda}$ is
Gaussian, it may be performed explicitly, leading to the saddle
point $H_{\mu\nu\lambda} = - \epsilon_{\mu\nu\lambda\rho} D^\rho
a$, where the covariant derivative
\be
    D_\mu a = \partial_\mu a - 4k \, A_\mu \,,
\ee
is invariant under the $U(1)$ transformations
\be
    \delta A_\mu = \partial_\mu \omega \qquad\hbox{and} \qquad
    \delta a = 4k \omega \,.
\ee
The resulting dual action for $a$ then becomes:
\be \label{GSaction3}
    \tilde S_{\scriptscriptstyle GS} = - \int \exd^4x \; \left[
    \frac12 \, D_\mu a D^\mu a + \frac{\kappa}{4} \, a\,
    \epsilon^{\mu\nu\lambda\rho} F_{\mu\nu} F_{\lambda\rho}
    \right] \,,
\ee
which again reproduces the proper anomalous $U(1)$ transformation.

Within a 4D $N=1$ supersymmetric context the `constants' $k$ and
$\kappa$ typically depend on various moduli fields, but the above
arguments carry through basically unchanged. In this case the
scalar $a$ resides within a complex chiral scalar multiplet, $\rho
= \kappa(z + 2i a)$, and so the second term in the action
\pref{GSaction3} requires $\rho$ to appear linearly in the
holomorphic gauge kinetic function: $f = \rho + \cdots$, and so
\be
    S_g = - \frac14 \int \exd^4x \; \Bigl[ \sqrt{-g} \; \kappa\,
    z\, F_{\mu\nu} F^{\mu\nu} + \kappa a
    \epsilon^{\mu\nu\lambda\rho} F_{\mu\nu} F_{\lambda\rho} \Bigr]
    \,.
\ee

By contrast, the kinetic term for $a$ and for the charged matter
fields, $\varphi$, arise from the K\"ahler function $K = K_1(\rho
+ \overline\rho + c V) + K_2(\overline\varphi e^{qV}, \varphi)$,
where $V$ denotes the $U(1)$ gauge multiplet and $c = -4\kappa k =
-Aq/(4\pi^2)$ is required in order to ensure that $\partial_\mu a$
and $A_\mu$ only appear through the invariant combination $D_\mu a
= \partial_\mu a - 4k A_\mu$. The K\"ahler function is also the
source of the $D$-term contributions
\be
    \left[\left(  c \, \frac{\partial K_1}{\partial\rho} + q\overline\varphi
    \frac{\partial K_2}{\partial \overline\varphi} \right)_{V=0} \,
    V \right]_D = q\left( - \frac{A}{4\pi^2} \,
    \frac{\partial K_1}{\partial\rho} + \overline\varphi
    \frac{\partial K_2}{\partial \overline\varphi} \right)_{V=0} D \,.
\ee
which show that the relative size of the two contributions is
independent of the sign of $q$. In particular, using $\kappa^2 K_1
= -3 \ln (\rho + \overline\rho)$ and the `minimal' choice, $K_2 =
\overline\varphi e^{qV} \varphi$, implies
\be
    D \propto q \left(\frac{3A}{4\pi^2 \kappa^2 (\rho + \overline\rho)}
    + \overline\varphi \varphi \right) \,,
\ee
in agreement with refs.~\cite{SSeq}. Clearly, in the absence of
any other $\varphi$-dependence, the $D$-term potential $V_D
\propto D^2$ is minimized by $\varphi = 0$ because of the
conditions that $\kappa^2(\rho + \overline\rho)/A$ is positive.

\end{document}